\documentclass[a4paper,twocolumn,
english,aps,prl,floatfix,showpacs]{revtex4}
\usepackage[latin1]{inputenc}
\usepackage{amsmath}
\usepackage{babel}
\usepackage{graphics}
\usepackage{amssymb}

\def\jap{J.\ Appl.\ Phys.\ }

\newcommand{\MnGa}{Mn$_{\rm Ga}$\ }
\newcommand{\up}{\uparrow}
\newcommand{\dn}{\downarrow}

\begin{document}


\title{On the nature of ferromagnetism in Ga$_{1-x}$Mn$_x$As semiconductors}

\author{Ant\^onio J. R. da Silva}
\affiliation{Instituto de F\'\i sica, Universidade de S\~ao Paulo, Cp 
66318, 05315-970, S\~ao Paulo SP, Brazil}
\author{A. Fazzio}
\affiliation{Instituto de F\'\i sica, Universidade de S\~ao Paulo, Cp 
66318, 05315-970, S\~ao Paulo SP, Brazil}
\author{Raimundo R. dos Santos}
\affiliation{Instituto de F\'\i sica, Universidade Federal do
Rio de Janeiro, Caixa Postal 68528, 21945-970 Rio de Janeiro RJ, Brazil}
\author{Luiz E. Oliveira}
\affiliation{Instituto de F\'{\i}sica, Unicamp, C.P. 6165,
13083-970 Campinas SP, Brazil}

\begin{abstract}
We have performed {\it ab initio} calculations within the density-functional
theory for Ga$_{1-x}$Mn$_x$As diluted semiconductors. Total energy results 
unambiguously show that a quasi-localized $\downarrow$ hole, with strong $p$-like 
character, surrounds the fully polarized Mn $\uparrow$ $d^5$-electrons. Calculations indicate that the
holes form an essentially dispersionless impurity band, thus rendering effective-mass descriptions
of hole states open to challenge. We obtain estimates both for
the $s = 1/2$ hole and $S = 5/2$ Mn exchange coupling, and for the
distance dependence of the effective Mn-Mn exchange interaction. Results demonstrate 
that the effective Mn-Mn coupling is always ferromagnetic, thus non-RKKY, and 
intermediated by the antiferromagnetic coupling of each Mn spin to the holes.
\end{abstract}

\date{Version 02.00 -- \today}


\maketitle

In the last few years, a considerable amount of work has been devoted to
the study of diluted magnetic semiconductors (DMS), since the possibility
of manipulating both the charge and spin degrees of freedom of carriers in
magnetic materials will change qualitatively the efficiency of spintronics
devices. The discovery of hole-induced ferromagnetism in p-type (In,Mn)As
systems \cite{Ohno92} was followed by the successful growth of
ferromagnetic (Ga,Mn)As alloys \cite{Ohno96}. A quantitative understanding
of the physics in these materials is therefore crucial, since
ferromagnetic III-V alloys may be readily combined into semiconductor
heterostruture systems, opening up a range of applications of
optoelectronic devices through the combination of quantum and magnetic
phenomena in these materials. However, in order to develop full-scale
applications, one needs to elucidate several issues in relation to these
systems. In the case of Ga$_{1-x}$Mn$_x$As semiconductors, it is well known
that Mn forms acceptors when in substitutional Ga lattice sites (Mn$_{\rm Ga}$).
One of the standing issues is related to the fact that the critical temperature and hole concentration, as a function of Mn
composition in Ga$_{1-x}$Mn$_x$As, are crucially dependent on the details of
growth conditions 
\cite{vanEsch97,Matsukura98,Ohno99a,Potashnik01,Edmonds02,Seong02,Asklund02,Yu02a,Yu02b,Potashnik02,Moriya03,dosSantos},
even in as-grown samples; note, in particular, that for 
$x\simeq 5\%$, $T_c$ ranges
from $\sim 30$ K to 110 K, with the highest temperature \cite{Matsukura98}
of 110 K only being reproduced by other groups after post-growth annealing \cite{Potashnik01,Potashnik02,Hayashi01}.
The measured hole concentration
$p$ as a function of $x$ similarly reveals
discrepancies both in trends and in order of magnitude; note that
different experimental techniques (Hall resistance \cite{Potashnik01} and Raman scattering \cite{Seong02})
applied to the same $x=0.083$ sample
yield values of $p$ differing by as much as a factor of 10. 

From the theoretical point of view, different models have been proposed to
describe the electronic and magnetic properties of (Ga,Mn)As. The
generally accepted view is that a given Mn ion interacts with the holes
via a local antiferromagnetic Kondo-like exchange coupling $J_{hd}\equiv
N\beta\cdot (a^3/4)$ ($a$ is the GaAs lattice parameter) between their magnetic
moments \cite{Ohno99a,Dietl97,Ohno01}; this interaction is thought
to lead to a polarization of the hole subsystem, which would then give
rise to an effective ferromagnetic coupling between the Mn moments.  
Early attempts treated the holes within an $sp$ parabolic-band
effective-mass approximation; this approach is in conflict with recent
photoemission studies \cite{Asklund02} and infrared measurements
\cite{Singley02}, which indicate that the holes form a relatively flat impurity band at
the Fermi energy, instead of residing in an unaltered GaAs valence band.
In particular, infrared spectroscopic measurements estimated the hole
effective mass within the range $0.7\, m_e < m^* < 15\, m_e$ for an
$x=0.52$ sample, and even larger values at higher dopings
\cite{Singley02}. As a consequence, any treatment of holes within an
effective-mass approximation, including attempts in the direction of
incorporating aspects such as a Kohn-Luttinger treatment of the valence
states \cite{Dietl01b,Abolfath01}, are quite clearly open to question.  
The same applies to estimates of $N\beta\sim 1.5 - 3$ eV
\cite{Matsukura98,Ohno01}, which are obtained from resistivity fits to
experimental data resorting to a hole effective mass $m^*=0.5\, m_e$.

Therefore, it is fundamental to have a highly accurate microscopic description of
the electronic and magnetic properties of Ga$_{1-x}$Mn$_x$As DMS. Here we perform a detailed {\it ab
initio} study of the physical origin of the observed Mn-Mn ferromagnetic coupling.
Our total energy results provide
unambiguous evidence that (i) the impurity states are essentially localized
and hence almost dispersionless, thus rendering any effective-mass picture inapplicable, and
(ii) the effective coupling is always
ferromagnetic (thus non-RKKY), intermediated by an antiferromagnetic
coupling of each Mn spin to the holes. In addition, we
have also provided reliable values for the Mn-hole exchange coupling, as
well as for the anisotropy- and direction-dependence of the effective
(i.e., if the hole degrees of freedom were integrated out) coupling
between Mn spins.

We have performed total energy calculations based on the
density-functional theory (DFT) within the generalized-gradient
approximation (GGA) for the exchange-correlation potential, with the electron-ion interactions
described using ultrasoft pseudopotentials \cite{Vanderbilt90}. A plane wave expansion up to 230 eV as implemented
in the VASP code \cite{Kresse93} was used, together with a 128-atom and a 250-
atom fcc supercell and the $L$-points for the Brillouin zone sampling. The
positions of all atoms in the supercell were relaxed until all the forces
components were smaller than 0.02 eV/\AA. We have also checked for spin-orbit effects 
through the projector augmented-wave (PAW) method \cite {Kresse99} and found that they 
may be safely neglected.  

Let us first consider the case of a single isolated \MnGa acceptor. From total energy calculations for neutral (Mn$_{\rm Ga}$)$^0$
and negatively charged (Mn$_{\rm Ga}$)$^-$ defects, with a 250-atom supercell, we obtain a localized acceptor level lying at 0.1 eV above the top of the valence band, in good agreement with the 0.11 eV experimental value \cite {Mnlevel}. Moreover, the ground state of the \MnGa defect is consistent with the picture of a $\dn$ hole interacting antiferromagnetically with the $\up$ ${\rm S} = 5/2$ spin of the $d^5$-configuration at the Mn site. The robustness of this state is illustrated by the fact that the ferromagnetic configuration with an $\up$ hole lies $\simeq$ 0.25 eV above the antiferromagnetic one. Figure \ref{singleMn}(a) shows the difference $m({\bf r}) \equiv \rho_{\up}({\bf r}) - \rho_{\dn}({\bf r})$ for the \MnGa defect, where $\rho_\sigma$ is the total charge density in the $\sigma$-polarized channel. We note that near the \MnGa acceptor, the local magnetization has a strong $\up$ $d$-like character, due essentially to the valence-band resonant $d^5$ electrons, whereas as one approaches its As neighbors, the character changes to $\downarrow$ p-like. The signature of the above mentioned antiferromagnetic interaction consists of a sign change in $m({\bf r})$ as one moves from the Mn site to any of its neighboring As. In order to probe the extent of the electronic charge associated with the \MnGa defect, the difference between $\rho_{\up}({\bf r}) + \rho_{\dn}({\bf r})$ calculated for the \MnGa state and the GaAs host is depicted in Figure \ref{singleMn}(b). One clearly notes the localized nature of the defect, as the charge difference is essentially confined within the region surrounding the Mn site. 

The origin of ferromagnetism in diluted Ga$_{1-x}$Mn$_x$As semiconductors may be elucidated by focusing on
interacting \MnGa substitutional defects in a 128-atom supercell, considering both a ferromagnetic as well as an antiferromagnetic allignment between the Mn spins. We have performed calculations for two Mn substitutional atoms in configurations corresponding to all inequivalent positions within the supercell, i.e., Mn-Mn distances varying from 4.06 
\AA\ up to 11.48 \AA. Our total energy results yield an unanbiguous Mn-Mn ferromagnetic ground state in all cases \cite {so}. Figure \ref{doubleMn} shows the net magnetization $m({\bf r})$ isosurfaces for two Mn defects in nearest-neighbor and next-nearest neighbor positions (for the Ga sublattice), with parallel and anti-parallel Mn spins.
In Fig. \ref{doubleMn}, the Mn atoms are at the center of the spherical-like regions of $m({\bf r})$, whereas the $p$-like
regions are always centered on As atoms. Note that, irrespective of the relative orientation of the Mn-Mn spins,
the antiferromagnetic coupling between the Mn and hole spins is always maintained. In the Mn-Mn antiferromagnetic case, this leads to the appearence
of nodes in $m({\bf r})$, which contributes to increase its energy relative to the ferromagnetic state. Another important feature illustrated in Fig. \ref{doubleMn} is the fact that, again, the
Mn-Mn defect is essentially localized, although the magnetization density clearly spreads out from one \MnGa site
to the other. The picture that emerges is that of a cloud of $\downarrow$ holes surrounding the substitutional Mn, with the distribution of quasi-localized holes giving rise to a nearly dispersionless impurity band, so that a description via the effective-mass approximation would be inappropriate.

From the difference between total energies of the ferro- and antiferromagnetic configurations of
the $s=1/2$ hole and the $S=5/2$ Mn spins, assuming an interaction of the type
$N\beta \, {\bf s}\cdot{\bf S}$, and taking into account the contribution due to images in neighboring
supercells, one may calculate the strength of the Kondo-like antiferromagnetic
exchange coupling $N\beta \sim +\ 0.1$ eV. This result contrasts with the previous estimates of 
$N\beta\sim 1.5 - 3$ eV \cite{Matsukura98,Ohno01}. We note that these larger values are open to question, as they are obtained via fittings, to experimental data, of results based on the effective-mass approximation with a hole effective mass $m^*=0.5\, m_e$. We have also estimated, from total energy calculations, the effective exchange coupling between pairs of $S=5/2$ Mn spins, $J_{Mn-Mn}$, as a function of the Mn-Mn distance, for all inequivalent pair positions within the supercell (for a Mn-Mn interaction modelled through $J_{Mn-Mn}\, {\bf S}_{{Mn_i}}\cdot{\bf S}_{{Mn_j}}$). In Fig. \ref{exchange-J} we display the theoretical predictions for $J_{Mn-Mn}$. The results clearly  show that the coupling between the Mn spins is always ferromagnetic, irrespective of their relative distance. As it is well known that the bare coupling between two Mn spins should be antiferromagnetic, one concludes that the resulting Mn-Mn ferromagnetic effective coupling, in Ga$_{1-x}$Mn$_x$As, is essentially intermediated by the antiferromagnetic coupling of each Mn spin to the quasi-localized holes. Also, the observed non-monotonic behavior of $J_{Mn-Mn} ({\bf r})$ should be attributed to the anisotropic character of the effective interaction, as it may be inferred from Fig. \ref{doubleMn}. The inescapable conclusion from Fig. \ref{exchange-J} is that an  RKKY description for the Mn-Mn interaction in Ga$_{1-x}$Mn$_x$As is ruled out. Here we note that this result contrasts with the RKKY coupling obtained by Zhao {\it et al.} \cite{Zhao03} in the case of Mn$_x$Ge$_{1-x}$.

The data in Fig.\ \ref{exchange-J} can be used to estimate the critical temperature at a given Mn concentration, which, within a simple mean-field theory is given by 
\begin{equation}
k_B T_c=\frac{S(S+1)}{3}\,|J_0|,
\label{tc}
\end{equation} 
with
\begin{equation}
J_0=\sum_{\bf r} J ({\bf r})= \left( \frac{\overline{z_1}}{2}J_1+\frac{\overline{z_2}}{2}J_2+\frac{\overline{z_3}}{2}J_3+\ldots\right),
\label{J0}
\end{equation}
where $J_1,\ J_2,\ldots$ stand for first-, second-,..., -neighbor interactions, the values of which are given 
in Fig.\ \ref{exchange-J}, and the $\overline{z_i}$ are the corresponding configurationally-averaged coordination numbers in the Ga sublattice. If we take $\overline{z_i}=fz_i$, with $f$ being the fraction of \emph{active} Mn sites, which is roughly $x/3$, one obtains $T_c=4\times 10^3 x$ K, in qualitative agreement with the low-density behavior observed experimentally \cite{Matsukura98}.

One should mention
that the discrepancies between different experimental data
indicate that the magnetic, structural, and electronic properties of
as-grown Ga$_{1-x}$Mn$_x$As alloys are extremely sensitive to the actual
molecular-beam epitaxy (MBE) growth conditions, such as, for example,
growth temperature and beam flux ratios. In fact, this is to be expected,
since the holes provided by the Mn$_{\rm Ga}$ acceptors may be compensated
by defects such as arsenic antisites (As$_{\rm Ga}$) donors, Mn
interstitials (Mn$_I$), Mn$_I$-Mn$_{\rm Ga}$ pairs, MnAs complexes, etc.
Of course, the situation with respect to annealed Ga$_{1-x}$Mn$_x$As samples is rather
more complicated, and a proper analysis of experimental measurements
performed in after-annealing samples must involve a realistic
modelling of diffusion processes involving several defects, formation of random
precipitates, clustering effects, etc. Therefore, we emphasize that
a proper understanding of the physics of Ga$_{1-x}$Mn$_x$As alloys must
involve a microscopic description of the effect of different defects on
their electronic and magnetic properties.

In conclusion, we have provided a detailed {\it ab initio} study of the  
physical origin of the Mn-Mn ferromagnetic coupling, by considering isolated  
\MnGa defects, as well as two substitutional Mn per supercell, in various 
relative positions. Our total energy calculations provide 
unambiguous evidence that the effective coupling is always
ferromagnetic, and thus non-RKKY, and intermediated by an antiferromagnetic
coupling of each Mn spin to the quasi-localized holes. 

Partial financial support by the Brazilian Agencies CNPq, CENAPAD-Campinas, Rede Nacional de Materiais Nanoestruturados/CNPq,
FAPESP, FAPERJ, and Millenium Institute for Nanosciences/MCT is gratefully acknowledged.

\newpage


\begin{figure} [h]
\caption{Isosurfaces for (a) the net local magnetization $m({\bf r}) = \rho_{\up}({\bf r}) - \rho_{\dn}({\bf r})$ for the \MnGa defect, and (b) the difference between $\rho_{\up}({\bf r}) + \rho_{\dn}({\bf r})$ calculated for the \MnGa ground state and the GaAs host. The green surface corresponds to a value of 
$+ \ 0.004 \ e/\text{\AA}^3$, and the blue surface to $- \ 0.004 \ e/\text{\AA}^3$; for comparison, note that a uniform charge density of 1 electron/unit cell in GaAs corresponds to $+ \ 0.022 \ e/\text{\AA}^3$, with $e$ being the electron charge. The black (red) spheres denote the Ga (As) atoms.}
\label{singleMn}
\end{figure}

\begin{figure}[h]
\caption{Isosurfaces for the net local magnetization $m({\bf r}) =
\rho_{\up}({\bf r}) - \rho_{\dn}({\bf r})$ in the case of two \MnGa
defects; the color code and isosurface values are the same as in Fig.\ 1.
In (a) and (b) [(c) and (d)] the two Mn are nearest neighbors
[next-nearest neighbors] with their $S=5/2$ spins alligned parallel and
anti-parallel, respectively.}
\label{doubleMn}
\end{figure}

\begin{figure}[h]
\caption{Exchange coupling $J_{Mn-Mn}$ between Mn atoms in Ga$_{1-x}$Mn$_x$As alloys, versus the Mn-Mn distance. Open dots are our calculated results whereas the full line is a guide to the eye.}
\label{exchange-J}
\end{figure}



\begin{thebibliography}{99}

\bibitem{Ohno92}  H.\ Ohno 
{\it et al.,}
\prl {\bf 68}, 2664 (1992).

\bibitem{Ohno96} H.\ Ohno
{\it et al.,}
\apl {\bf 69}, 363 (1996).

\bibitem{vanEsch97}  A.\ Van Esch
{\it et al.,}
\prb {\bf 56}, 13103 (1997).

\bibitem{Matsukura98} F.\ Matsukura
{\it et al.,}
\prb {\bf 57}, R2037 (1998).

\bibitem{Ohno99a}  H.\ Ohno, J.\ Magn.\ Magn.\ Mater.\ {\bf 200}, 110 (1999).

\bibitem{Potashnik01} S.\ J.\ Potashnik
{\it et al.,}
\apl {\bf 79}, 1495 (2001).

\bibitem{Edmonds02} K.\ W.\ Edmonds
{\it et al.,}
\apl {\bf 81} 3010 (2002). 

\bibitem{Seong02} M.\ J.\ Seong
{\it et al.,}
\prb {\bf 66}, 033202 (2002).

\bibitem{Asklund02} H.\ Asklund
{\it et al.,}
\prb {\bf 66}, 115319 (2002).

\bibitem{Yu02a} K.\ M.\ Yu
{\it et al.,}
\prb {\bf 65} R201303 (2002). 

\bibitem{Yu02b} K.\ M.\ Yu
{\it et al.,}
\apl {\bf 81} 844 (2002). 

\bibitem{Potashnik02} S.\ J.\ Potashnik
{\it et al.,}
\prb {\bf 66}, 012408 (2002).

\bibitem{Moriya03} R.\ Moriya and H.\ Munekata, \jap {\bf 93}, 4603 (2003).

\bibitem{dosSantos} R.\ R.\ dos Santos {\it et al.,}
J. Phys.:Cond. Matter {\bf 14}, 3751 (2002); \jap {\bf 93}, 1845 (2003).


\bibitem{Hayashi01} T.\ Hayashi
{\it et al.,}
\apl {\bf 78}, 1691 (2001).

\bibitem{Dietl97} T.\ Dietl
{\it et al.,}
\prb {\bf 55}, R3347 (1997).

\bibitem{Ohno01} H.\ Ohno and F.\ Matsukura, Solid State Commun.\ {\bf
117}, 179 (2001).

\bibitem{Singley02} E.\ J.\ Singley
{\it et al.,}
\prl {\bf 89}, 097203 (2002).

\bibitem{Dietl01b}  T.\ Dietl
{\it et al.,}
\prb {\bf 63}, 195205 (2001).

\bibitem{Abolfath01}  M.\ Abolfath 
{\it et al.,}
\prb {\bf 63}, 054418 (2001).

\bibitem{Vanderbilt90} D.\ Vanderbilt, \prb {\bf 41}, 7892 (1990).

\bibitem{Kresse93} G.\ Kresse and J.\ Hafner, \prb {\bf 47}, R558 (1993);
G.\ Kresse and J.\ Furthm\"uller, \prb {\bf 54}, 11169
(1996).

\bibitem{Kresse99} G. Kresse and D. Joubert, \prb {\bf 59}, 1758 (1999).

\bibitem{Mnlevel} R. A. Chapman and W. G. Hutchinson, \prl {\bf 18}, 443 (1967); J. Schneider {\it et al.,} \prl {\bf 59}, 240 (1987); M. Linnarsson {\it et al.,} \prb {\bf 55}, 6938 (1997).

\bibitem{so} In the case of two nearest-neighbor Mn defects in a 128-atom supercell, we have considered spin-orbit effects within the PAW method
({\it cf.} Ref. \onlinecite{Kresse99}) as implemented in the VASP code, and found a change from 0.29 eV to 0.24 eV in the total energy difference between the excited antiferromagnetic and ground state ferromagnetic Mn-spin alignements. As this change is not substantial, we therefore choose to ignore spin-orbit effects in the total energy calculations presented in this work. 

\bibitem{Zhao03} Y-J. Zhao {\it et al.,} \prl {\bf 90}, 047204 (2003).

\end{thebibliography}
\end{document}